\documentclass[11pt]{article}
\usepackage{amsmath,amssymb,color}
\usepackage{amsfonts}
\usepackage{graphicx}

\newcommand{\hoch}[1]{$\, ^{#1}$}
\newcommand\ov{\over}
\newcommand\p{\partial}
\newcommand{\be}{\begin{equation}}
\newcommand{\ee}{\end{equation}}
\newcommand{\bea}{\begin{eqnarray}}
\newcommand{\eea}{\end{eqnarray}}

\begin{document}

\begin{flushright}
\hfill{ \
SHUTP-11-19\ \ \ \ }

\end{flushright}

\vspace{25pt}
\begin{center}
{\large {\bf Transport Coefficients for Holographic Hydrodynamics at Finite Energy Scale }}
\vspace{15pt}

  Xian-Hui Ge{\hoch{1}'\hoch{2}} , Hong-Qiang Leng\hoch{1}, Li Qing Fang\hoch{1} and Guo-Hong Yang\hoch{1}

\vspace{10pt}
\hoch{1}{\it Department of Physics, Shanghai University, Shanghai 200444, China}
\vspace{10pt}

\vspace{10pt}
\hoch{2}{\it State Key Laboratory of Theoretical Physics,Institute of Theoretical Physics,Chinese Academy of Sciences,Beijing 100190,China}
\vspace{10pt}

\underline{ABSTRACT}
\end{center}

  We investigate the relations between black hole thermodynamics and holographic transport coefficients in this paper. The formulae for DC conductivity and diffusion coefficient are verified for electrically single-charged black holes. We examine the correctness of the proposed expressions by taking
 charged dilatonic and single-charged STU black holes as two concrete examples, and compute the flows of conductivity and diffusion coefficient by solving the linear order perturbation equations.
 We then check the consistence by evaluating the Brown-York tensor at a finite radial position.  Finally, we find that the retarded Green functions for the shear modes can be expressed easily in terms of black hole thermodynamic
 quantities and transport coefficients.

\vspace{15pt}

\thispagestyle{empty}

\newpage

    \section{Introduction}
The experiments at the Relativistic Heavy Ion Collider
(RHIC) and at the Large Hadron Collider (LHC)  show that the quark-gluon plasma
(QGP) does not behave as a weakly coupled gas of quarks and gluons, but rather as a
strongly coupled fluid. This places limitations on the applicability of perturbative
methods. The AdS/CFT correspondence provides a powerful tool for studying the dynamics of strongly coupled quantum field theories\cite{Maldacena:1997re,Gubser:1998bc,Witten:1998qj,Susskind}.  Moreover, the result of RHIC experiment on
the viscosity/entropy ratio turns out to be in favor of the
prediction of AdS/CFT ~\cite{pss0,kss0,bl} and some attempt has been
made to map the entire process of RHIC experiment in terms of
gravity dual \cite{ssz}.

Aimed to develop a model independent  theory of the hydrodynamics,
the membrane paradigm and a holographic version of Wilsonian Renormalization Group (hWRG) have been
proposed to describe strongly coupled field theories with a finite cut-off \cite{Joe.RG,Andy.RG,Sin:2011yh,Hong.RG,Hong.Membrane,sz,ge3,melgar,park,oh}.
 The radial flow in the bulk geometry can be regarded as the renormalization group flow of the boundary and the radial direction marks the energy scale of the boundary field theory\cite{peet,ak,al,gi,di,ba,free,boer,boer1,li} and it was found in \cite{nojiri} that the holographic renormalization group  may also lead to instable AdS background. For neutral black hole duals, some universal transport coefficients of the generic boundary theory can be expressed in terms of geometric quantities evaluated at the horizon\cite{Hong.Membrane}. It was later proved that the expressions given by the membrane paradigm are universal for the various neutral black holes.

On the other side, we know that the calculation of the linear response functions (i.e. the retarded Green functions) of a strongly coupled system is very important, but obviously a tough task. Even for the transport coefficients of translational-invariant charged
black holes, the complete solutions of the shear modes and sound modes are coupled together and hard to be solved \cite{Andy.RG,sz}. At finite momentum, it has been found that the equation of motion for the shear modes of the charged black hole in higher derivative gravity turned out to be impossible to decouple\cite{ge3}, let alone the sound modes.  However, compared with the tedious calculation of the linear response, the black hole thermodynamics seems simple and concise.
The black hole no-hair theorem asserts that  all black hole solutions in general relativity can be completely characterized by only three externally observable classical parameters: mass, electric charge, and angular momentum.

A natural question is that why the linear response functions of  a two parameters (mass and charge) system is so complicated to deal with and do we have a more general and powerful method to deal with the linear type perturbation of charged black holes ? We are going to present a positive answer.

 In \cite{Andy.RG} and \cite{sz}, the membrane paradigm and the hWRG approach were utilized to investigate the transport coefficients of the charged AdS black hole at an arbitrary cut off surface between the horizon and the boundary. In particular, it was found in \cite{sz} that the diffusion coefficient for Reissner-
Nordstrom-Anti de Sitter(RN-AdS) black holes \cite{gks} can be computed in the scaling limit without
explicit decoupling procedure.  In \cite{ge3}, they applied the hWRG approach  to the Einstein-Maxwell-Gauss-Bonnet theory  and utilized a formula for DC conductivity at an arbitrary cut off surface for charged black holes
\be\label{sig}
\sigma_x(u_c)= \frac{1}{q^2(u)
}\sqrt{\frac{-g}{g_{uu}g_{tt}}}g^{xx}\bigg|_{u=1}\bigg(\frac{T_c s_c}{\varepsilon+P}\bigg)^2,
\ee
where $q^2(u)$ denotes the gauge coupling, $g$ and $g_{\mu\nu}$  the determinant and metric component, and $s$, $T_c$, $\epsilon$ and $P$ are the entropy density, temperature, energy density and pressure evaluated at the cut-off surface, respectively. The formula (\ref{sig}) reflects the fact the conductivity varies with the sliding membrane,  and there is a RG flow between the horizon and the boundary. For neutral black holes, the conductivity (\ref{sig}) recovers the result given by Iqbal and Liu\cite{Hong.Membrane} because of the Euler relation $\varepsilon+P=T  s $.

Our logic of this paper is as follows:  When a system is perturbed slightly, its response will be linear in the perturbation and this regime is called the linear response regime. Although the system is in a non-equilibrium state  of which all characteristics can be inferred from the properties of its equilibrium state. Because all the scalar, shear and sound modes are linear response to small perturbations to the  black hole thermodynamic equilibrium state, all the transport coefficients can be determined by the black hole thermodynamic variables in its equilibrium state.

The purpose of this paper is to verify the unified form of  DC conductivity and  diffusion constant for translational-invariant hydrodynamics with a chemical potential, as a first step towards general formulae of transport coefficients of anisotropic and inhomogeneous hydrodynamics. The formula (\ref{sig}) holds for most single charged black holes\footnote{For multi-charge black holes, the conductivity should be a matrix and has been investigated in dual rotating D3,M2 and M5 brane by Jain in \cite{jainmuti}.} and we can rewrite  Eq.(\ref{sig}) in terms of the metric components. So, if somebody knows the black hole solution, the DC conductivity can be computed directly by using
  \be\label{sigmadc}
\sigma_{DC}(u_c)=\frac{\Omega(1)}{\digamma^2(u_c)},
\ee
\be
\Omega(1)=\bigg(\frac{g^{(5d-2)/4}_{xx}g'_{tt}}{q(u)\sqrt{-g}}\bigg)^2\bigg|_{u=1}, \nonumber
\ee
\be
\digamma(u_c)=\sqrt{-g}g^{uu}\bigg(\ln {\frac{-g_{tt}}{g_{xx}}}\bigg)'\bigg|_{u=u_c},  \nonumber
\ee
where $(d+2)$ is the bulk spacetime dimensions, and the prime $'$ hereafter denotes the derivative with respect to $u$.

As to the diffusion coefficient, we  also conjecture the following formula for holographic hydrodynamics with a chemical potential
\bea
&&\bar D(u_c)=\frac{1}{4\pi}\frac{\Xi(1)}{\digamma(u_c)},\\
&&\Xi(1)=\frac{-g^d_{xx}g'_{tt}}{\sqrt{-g}}\bigg|_{u=1}, \nonumber
\eea
where$\sqrt{-g}$ is determinant of the metric.

After that, we will  prove that $\digamma(u)\propto \frac{\epsilon+P}{T s}$ actually is the solution of the transverse mode perturbation equation of the gauge field in the zero frequency and zero momentum limit.
As it was noticed that the $U(1)$ charge makes the hydrodynamics analysis complicated \cite{sz,ge3} because vector modes of gravitational fluctuation mix with transverse Maxwell modes.

In fact, the formula (1) has been proposed  in the previous paper \cite{jain1008}.  In that paper, the author investigated the electrical conductivity and thermal conductivity by using the proposed formulae for  RN black hole and charged Lifshitz black hole. However,  the author did not mention how to calculate the diffusion coefficient. In this work, we are going to calculate the transport coefficients for charged dilatonic black hole and R-charged black hole.  In addition, the formulae of diffusion coefficient and retarded Green functions will be proposed in a generalized way.

Before going on, let us summarize the new features and the main result of this paper:
\begin{itemize}
\item  The unified form of the diffusion coefficient and the retarded Green functions are proposed for the first time.
\item By evaluating the black hole thermodynamic quantities as a function of radial coordinate, we can write down the transport coefficients and the retarded Green functions in terms of black hole metric line-element and black hole thermodynamic variables. This result implies that there exists deep
connection between black hole thermodynamics and the linear response functions.
\item Our work can be regarded as a first step towards easy computation of the transport coefficients with respect to the sound modes and holographic lattice in which partial differential equations are involved.

\end{itemize}
The contents of this paper are organized as follows: In section 2, we will consider the shear modes of charged dilatonic black holes. At first part of this section, we will review on the black hole geometry and thermodynamics. By using the RG flow approach
developed in \cite{Andy.RG}, we will compute the conductivity and diffusion coefficients at a  cut-off surface. Then we will provide a consistent check on the result by using the black hole thermodynamic relation and the Brown-York tensor.
The unified form of the retarded Green functions  evaluated on the boundary related to the shear modes will be presented in the appendix.
In section 3, we will work on the single-charged STU black holes. The conclusion will be presented in the last section.

\section{Charged Dilatonic Black Hole}
In this section, we study the transport coefficients and the RG flow of holographic hydrodynamics for the charge dilatonic black holes. We will first review on the black hole geometry and thermodynamics.  Then,
we will compute the transport coefficients.
\subsection{Backgrounds and Thermodynamics}
We start by introducing the following action for charged dilatonic black hole in $AdS_5$
with mixing dilatonic field and $U(1)$ gauge field \cite{Gubser:0911}
 \be
    {\cal L} = {1 \over 2\kappa^2} \left[ R - {1 \over 4} e^{4\alpha} F_{\mu\nu}^2 -
    12 (\partial_\mu\alpha)^2 + {1 \over L^2} (8e^{2\alpha} +
      4e^{-4\alpha}) \right] \,   ,  \label{chargedilatonaction}
\ee
 where $L$ is the radius of the $AdS$ space and $ \alpha $ plays the role of a dilaton with respect to the  radial coordinate $r$.
 We denote the gravitational constant as $\kappa^2=8 \pi G_{5}$ and the
 $U(1)$ gauge field strength is given by $F_{\mu\nu}(x)=\partial_\mu A_\nu(x)-\partial_\nu A_\mu (x)$ .

 In this paper, we will introduce a dimensionless coordinate  $ u=r_0/r$ for simplicity .
 The spatially uniform, electrically charged solution for this action  \cite{Gubser:0911}  can be obtained as follows
\be
    ds^2= e^{2A} (-h \, dt^2 + d\vec{x}^2) + {e^{2B}r_0^2 \over h u^4} du^2 ~~~ , ~~~
    A_\mu dx^\mu = \phi dt  \qquad~~~~~~~~~   \label{Onechargeexpress}\
\ee
where
 \be
       A =\log {r_0 \over u L} + {1 \over 3} \log\left( 1 + {Q^2 u^2\over r_0^2} \right), \qquad   B =-\log {r_0 \over u L} - {2 \over 3} \log\left( 1 + {Q^2 u^2\over r_0^2} \right) \nonumber \\
 \ee
 \be
        h = 1 - {u^4 m L^2 \over (r_0^2 + Q^2 u^2)^2}, \qquad  \phi = {Q u^2\sqrt{2m} \over r_0^2 + Q^2 u^2}-{Q\sqrt{2m} \over r_0^2 + Q^2}, \qquad \alpha = {1 \over 6} \log\left( 1 + {Q^2 u^2\over r_0^2} \right).
        \nonumber \\
 \ee
 The horizon of the black hole locates at
 \be
       r_0=\sqrt{L\sqrt{m}-Q^2}, \nonumber \\
 \ee
where $m$ is a constant  related to the mass  with the form
\be
m=\frac{(Q^2 +r_0^2)^2}{L^2} .  \label{mass} \\
\ee
The black hole is extremal if $r_0 = 0$, which implies $m=Q^4 / L^2 $.

The equation of motion for the gauge field $A_{\mu}(x)$ is given by
\be
\partial_\nu {(\sqrt{-g} e^{4 \alpha}F^{\mu\nu})}=0 . \label{maxwell}
\ee
The Einstein equation is written as
\be
R_{\mu\nu}-\frac{1}{2} g_{\mu\nu}R=G_{\mu\nu}+T_{\mu\nu} , \label{einsteine}
\ee
where
\bea
  && G_{\mu\nu}=\frac{1}{2} e^{4\alpha} F_{\mu\rho}F_{\nu}^{\rho}-\frac{1}{8} e^{4\alpha}  g_{\mu\nu}F^2 ,  \nonumber \\
   &&T_{\mu\nu}=12(\partial_\mu {\alpha} {\partial_\nu }{\alpha})-6 g_{\mu\nu} (\partial{\alpha})^2+
      \frac{1}{2 L^2} g_{\mu\nu} (8 e^{2\alpha}+4 e^{-4\alpha}) . \nonumber
\eea
The equation of motion for the scalar field is
\be
  -\frac{3}{4} \nabla^2 \alpha+\frac{1}{L^2} (e^{2 \alpha}-e^{-4 \alpha})=\frac{1}{16} e^{4\alpha} F^2.
\ee
The Hawking temperature yields
\be
     T=\frac{h'u^2}{4 \pi r_0}e^{A-B}\bigg|_{u= 1}=\frac{r_0}{\pi L^2}. \label{temperature}\
\ee
The volume density of Bekenstein-Hawking entropy  is given by
\be
    s=\frac{e^{3A}}{4G_5}\bigg|_{u= 1}={2r_0 \pi  \over \kappa^2 L^3}(Q^2+r_0^2). \label{entropy}\
\ee
According to thermodynamic relation
\be
    \varepsilon +P-s T-\mu \rho=0, \label{thefirst}
\ee
 and
the first law of thermodynamics
\be
  d\varepsilon=Tds+\mu d\rho , \label{energy}\
\ee
we can obtain the energy density and the pressure,  which can be expressed as
\be
 \varepsilon={3(Q^2+r_0^2)^2 \over 2\kappa^2 L^5}, \qquad  P={(Q^2+r_0^2)^2 \over 2\kappa^2 L^5}.
\ee
The charge density and the chemical potential are  \cite {Gubser:0911}
\be
  \rho = {Q \sqrt{2} \over \kappa^2 L^3}(Q^2+r_0^2), \qquad
  \mu ={\frac{\sqrt{2}Q}{L^2}} .
\ee
Furthermore, there are several  relations satisfied by the thermodynamic variables
 \bea
    &&s =\frac{2\pi^2L}{\kappa}T\sqrt{\frac{2}{3}L\varepsilon},\\
     &&s \mu=2\pi^2\rho T ,\\
     && \frac{\mu^2}{2}+T^2\pi^2=\frac{\sqrt{m}}{L^3},\\
      &&   \varepsilon=\frac{3\kappa^{\frac{2}{3}}}{2^{\frac{7}{3}}\pi^{\frac{4}{3}}L}(s^2+2\pi^2 \rho^2)^{\frac{2}{3}}.
 \eea
 The temperature and chemical potential can also be obtained as
 \be
      T=\bigg(\frac{\partial\varepsilon}{\partial s}\bigg)_\rho, \qquad
      \mu=\bigg(\frac{\partial\varepsilon}{\partial \rho}\bigg)_s.
 \ee
The susceptibility can be calculated which is given by
 \be
         \chi=\bigg(\frac{\partial\rho}{\partial\mu}\bigg)_T=\frac{(Q^2+r_0^2)}{\kappa^2 L}+\frac{2 Q^2}{\kappa^2L} . \label{sus}\
\ee
According to the definition of  the special heat
\be
C_{\mu}=T\bigg(\frac{\partial s}{\partial T}\bigg)\bigg|_{\mu}.
\ee
Especially, when the temperature of the charged dilatonic black hole becomes very low , there is a linear special heat which can be expressed as
\be
 C_{\mu}=\frac{2\pi^2Q^2}{\kappa ^2L} T.
\ee
\subsection{Perturbations and Transport Coefficients}

Considering the metric perturbations\cite{son1}
\be
   g_{\mu\nu}\rightarrow g_{\mu\nu}+h_{\mu\nu}, ~~~\quad ~~~A_\mu \rightarrow A_\mu+\delta A_\mu ,
\ee
to the background  $g_{\mu\nu}$ and $A_\mu$, one can use background metric  $g_{\mu\nu}$ and inverse metric $g^{\mu\nu}$ lower and raise  tensor indices.
 The inverse metric can be expressed as $g^{\mu\nu}=g^{(0)\mu\nu}-h^{\mu\nu} + {\cal O}(h^2)$

 We choose the momentum  along the $z$-direction  and $u$ as the radial direction which  describes
the energy scale in field theory.  In the gauge $A_u(x)=0$  and by using  Fourier decomposition
\bea
&&h_{\mu\nu}(t, z, u)=\int \frac{d^5k}{(2\pi)^5} {e}^{-i\omega t+ikz}h_{\mu\nu}(k, u),\\
&&A_\mu(t, z, u)=\int \frac{d^5k}{(2\pi)^5} {e}^{-i\omega t+ikz}A_\mu(k, u),
\eea
we will consider the scalar mode $h_{xy}(t, z, u)$  and shear mode $h_{tx}(t, z, u) $, $h_{zx}(t, z, u)$ and $A_x(t, z, u)$ in the following.
\subsubsection{Scalar Mode and Shear Viscosity}
From symmetry  analysis,
one can find that  off-diagonal perturbation $h^y_x $ decouples from all other perturbations.
We obtain an equation of motion for the scalar mode as
\be
\partial_\mu (\sqrt{-g} \partial^\mu h^y_x)=0\ .\label{tensormode}\
\ee
Following the sliding membrane argument \cite{Hong.Membrane}, we  define a cutoff dependent tensor response function
\be
G_{xy} (u_c, k) ={ -\sqrt{-g}g^{uu}\p_u h^y_x\ov 2\kappa^2 h^y_x(u_c, k)}\ .
\ee
We define the shear viscosity as
\be
\eta(u_c, k):={G_{xy} (u_c, k)\ov i\omega}.
\ee
At zero momentum limit, the flow equation is given by
\be
\label{etaflow}
{\p_{u_c}\eta(u_c, \omega)}=i\omega \left({2\kappa^2 \eta^2(u_c, \omega)\ov \sqrt{-g}g^{uu}}-\frac{\sqrt{-g}g^{tt}}{2\kappa^2}\right)\ .
\ee
The shear viscosity is requested to be
\be
\label{floweta}
\eta(r_0)=\frac{{r_0}^3}{2\kappa^2 L^3} (1+\frac{Q^2}{r_0^{2}}),
\ee
because of the horizon regularity. Because the entropy density was given in (\ref{entropy}), we can easily check the shear viscosity to entropy density ratio
\be
\frac{\eta}{s}=\frac{1}{4\pi}. \label{gener}
\ee
This result agrees with \cite{son1,son2,Son.Realgreenfunction} and obeys the KSS bound \cite{kss}.

There some debate on whether the shear viscosity flow depend on the position of the cut-off surface\cite{Andy.RG,sz}. From the fluid-gravity computation \cite{Andy.RG}, both the shear viscosity
and the entropy density depend on the cut-off position, but their ratio does not .

Actually, if we do not consider the quantum corrections to the geometry, the radial evolution of the total entropy  remains a constant in nature. Therefore, we can see that the entropy density must depend on the radial coordinate.
Isentropic evolution equation\cite{Andy.RG}
\be
\partial_uS=0  \label{equen}
\ee where $S=sV_3$ and $V_3=e^{3A}$ is the $u$-dependent volume. We can derive $u$-dependent entropy density from the thermodynamical relation (\ref{thefirst}). Multiplying (\ref{thefirst}) with the volume $V_3$ and considering
the derivative of the total entropy along the radial direction, we obtain
\be
\partial_uS=\frac{V_p}{16\pi Gr_0T_H}(h''+(4A'-B'+\frac{2}{u})h')e^{A-B}u^2-\frac{V_p}{T_H}T_{\mu\nu}\zeta^\mu\zeta^\nu.
\ee The right hand side of the above equation, is exactly a component of Einstein equation. Here $T_{\mu\nu}$ is the bulk matter stress tensor and $\zeta^\mu$ is any null vector tangent to the cutoff, $T_{\mu\nu}\zeta^\mu\zeta^\nu \geq0$ implies the null energy condition. So, we can turn around to state that the radial Einstein equation implies the isentropic character, the total entropy keeps a constant in everywhere (\ref{equen}).
In this sense, the entropy density is evaluated as
\be
s(u_c)=\frac{u^3_c}{4G_5}\frac{(Q^2+r_0^2)}{(Q^2u_c^2+r_0^2)}  \label{cutoffentr1}.
\ee
We will follow \cite{Andy.RG} and assume that  ${\eta}/{s}=1/4\pi$  (obeys the KSS bound \cite{kss}) at an arbitrary  cut-off surface \footnote{The KSS bound is violated by higher derivative gravity (see \cite{ge,kats,ge2,ge4} and references there in). The holographic RG flow in such gravity was done in \cite{ge3}. }.
Under such an assumption, we find the shear viscosity
\be\label{cutoffeta1}
\eta(u_c) =\frac{u_c^3}{2\kappa^2}\frac{(Q^2+r_0^2)}{(Q^2u_c^2+r_0^2)}.
\ee
\subsubsection{Shear Modes: $h_{tx}(t, z, u) $, $h_{zx}(t, z, u)$ and $A_x(t, z, u)$}
The linear  perturbative Einstein equation
can be read off from the $(t, x)$, $(u, x)$ and $(x, z)$ components, respectively
\bea
&&0={h^x_t}''+(\frac{2}{u}+4A'-B') {{h^x_t}'} -\frac{r_0^2}{u^4 h}e^{2B-2A}\Big(\omega k h^x_z+k^2 h^x_t\Big)+\phi' e^{4\alpha-2A} {A_x}', \label{eq_motion_v_001}\\
&&0=kh{h^x_z}'+\omega{h^x_t}'+\phi' e^{4\alpha-2A} \omega {A_x}, \label{eq_motion_v_002}\\
&&0={h^x_z}''+(\frac{2}{u}+4A'-B'+\frac{h'}{h}){h^x_z}'+\frac{r_0^2}{u^4 h^2}e^{2B-2A}\Bigl(\omega^2 h^x_z+\omega k h^x_t\Bigr),\label{eq_motion_v_003} \eea
where the prime denotes the derivative with respect to $u$.
Among these three equations for vector modes, there are only two independent equations, because (\ref{eq_motion_v_002}) is a constraint equation.
From the  Maxwell equation (\ref{maxwell}), the $x$-component gives
\begin{equation}
0
=
{A_x}''+(\frac{2}{u}+2A'-B'+4\alpha'+\frac{h'}{h}){A_x}'+\frac{r_0^2}{u^4 h^2}e^{2B-2A}\Bigl(\omega^2-k^2h\Bigr) {A_x}
+\frac{\phi'}{h}{h^x_t}'.
\label{eq_motion_v_005}
\end{equation}
One shall notice that in  the limit  of zero momentum , the equations for the metric and the gauge perturbations are completely decoupled.

For convenience of calculation, we would like to define ``current" and ``strength" for the vector field $h^{x}_t $ and $h^{x}_z$ as follows
\be
j^\mu = -{1\ov g_{\text{eff}}^2(u)}\sqrt{-g}f^{u\mu}\ ,\quad f_{\mu\nu} = \p_\mu h^x_\nu - \p_\nu h^x_\mu\ .
\ee
We also introduce the effective coupling  $q^2(u)$ and $g_{\text{eff}}^2(u)$
\be
G =\frac{\sqrt{-g}}{g^2_{\text{eff}}(u)}\ , \quad  \frac{1}{q^2(u)} = {e^{4\alpha}\ov 2\kappa^2}\ , \quad {1\ov g^2_{\text{eff}}(u)}= {g_{xx}\ov 2\kappa^2}\ .
\ee
The vector part off-shell action in $AdS_5$ charged dilatonic black hole can be written as follows
\be
S=\int d^{5}x\; \sqrt{-g} \left( -\frac{1}{4q^2(u) }  F^{\mu\nu}F_{\mu\nu}
        - \frac{1}{4g^2_{\text{eff}}(u)}  f^{\mu\nu}f_{\mu\nu}   +\frac{1}{q^2(u)}h^x_t A'_x F^{ut} \right)\ .
\ee
Note that $F_{\mu\nu}$ is the strength of the Maxwell fields  $A_\mu$ and should not be confused with the effective
strength of the shear modes of gravity $f_{\mu\nu}$.

One the other hand, we must introduce a new current related to $A_x$
\be
 j^\lambda=- \frac1{q^2(u)}\sqrt{-g} F^{ut} A_x\ .
\ee
Then we can define the shifted current
\be
    \tilde{j^t} \equiv  j^t +j^\lambda.\nonumber
\ee
In terms of the defined ``current" and ``strength", the equations of motion (\ref{eq_motion_v_001})-(\ref{eq_motion_v_003}) can be recast as
\bea
  &&\p_t\tilde{j^t} + \p_z j^z = 0,\label{heom0}\\
 && \p_u\tilde{j^t} + Gg^{tt}g^{zz}\p_zf_{zt} = 0, \label{heom1}\\
&&  \p_uj^z - Gg^{tt}g^{zz}\p_tf_{zt} = 0. \label{heom2}\
\eea
The Bianchi identity holds as
\be \label{heom3}
-{g_{uu}g_{zz}\ov G}\p_tj^z + {g_{uu}g_{tt}\ov G}\p_zj^t + \p_uf_{zt} = 0\ .
\ee
For the same reason,  we can define $j^x$ and $\tilde j^x$ as
 \be
j^x \equiv -\frac1{q^2(u)}\sqrt{-g} F^u{^x} \ , \quad
  \tilde{j}^x \equiv j^x +j^\rho,\quad  \ee
  where
\be
j^\rho=\frac1{q^2(u)}\sqrt{-g}  F^{ut}  h^x_t.
\ee
 The equation of motion for $A_x$ can be written as
\be
\label{Aeomx}
-\p_u \tilde{j^x} + \frac1{q^2(u)}\sqrt{-g}g^{xx}(g^{tt}\p_t F_{tx} + g^{zz}\p_z F_{zx}) = 0\ .
\ee
In the zero momentum limit,
 the equation of motion for $  A_x $   decouples from $h^x_t$
\be
\label{EOMAxD}
\p_u(\frac1{q^2(u)}\sqrt{-g}g^{uu}g^{xx}\p_u  A_x)- \frac1{q^4(u)}\sqrt{-g}g^2_{\text{eff}}(u)g^{uu}g^{tt}(\phi')^2 A_x -\frac1{q^2(u)}
\sqrt{-g}g^{xx}g^{tt}\omega^2 A_x =0\ .
\ee
The relevant on-shell action for $A_x$ at boundary $u=u_c$ can be written as
\be\label{AonshellS}
  S_{\text{on-shell}} =  \int_{u=u_c} \tilde j^x A_x\ .
\ee

\subsubsection{DC Electric Conductivity}
 By defining
\be
   \sigma_x=\frac{j^x}{E_x}=\frac{j^x}{F_{xt}} = {j^x \ov i\omega A_x},\  \label{sigmadef}\
\ee
we have
\be
\p_u\sigma_x = {\p_uj^x F_{xt} - j^x
\p_u(F_{xt})\ov F_{xt}^2}\ ,\quad \sigma_x{^2} = {(j^x)^2\ov(F_{xt})^2}\ .\label{sigma1}
\ee
So the flow equation (\ref{EOMAxD}) for
electric conductivity  can be rewritten as
\be\label{sigmaflow0}
{\p_u \sigma_x\ov i\omega} -
{q^2(u)\sigma_x^2\ov \sqrt{-g}g^{uu}g^{xx}} +{g^2_{\text{eff}}(u)\sqrt{-g}g^{uu}g^{tt}\phi'^2\ov \omega^2 q^4(u)}- \frac1{q^2(u)}\sqrt{-g}g^{xx}g^{tt} = 0\ .
\ee
The regularity condition at the horizon  requires
\be\label{SA}
\sigma_x(u=1)= \frac{1}{q^2(u)
}\sqrt{\frac{-g}{g_{uu}g_{tt}}}g^{xx}\bigg|_{u=1}=\frac{{r_0}}{2\kappa^2 L} \left(1+\frac{Q^2}{r_0^{2}}\right).
\ee
On the other hand, the DC conductivity can be evaluated by using the Kubo formula
\be
\sigma_{DC}=-\lim_{\omega\rightarrow 0}\frac{{ \rm Im} G^{R}_{x,x}(\omega,k=0)}{\omega}.
\ee
The retarded Green's function is given by \cite{son1,Son.Realgreenfunction}
\be
G^{R}_{x,x}(\omega,k=0)=-i\int dt d\textbf{x}e^{i\omega t}\theta(t)<[J_x(x),J_x(0)]>,
\ee
where $J_{\mu}$  is the conformal field theory (CFT) current dual to the bulk gauge field $A_{\mu}$.
It is convenient to define the radial momentum as \cite{ge3}
\be
J^x_k=\frac{1}{\kappa^2}K(u)A_x'(u,k).
\ee
The equation of motion (\ref{EOMAxD}) for $A_x$ can be rewritten as \cite{myers}
\be
\partial_uJ^x_k=\frac{1}{\kappa^2}L(u)A_x(u,k),
\ee
where
\be
K(u)=-\frac{\sqrt{-g}}{q^2(u)}g^{xx}g^{uu},~~~L(u)=\omega^2\frac{\sqrt{-g}}{q^2(u)}g^{xx}g^{tt}-\frac{g^2_{eff}\sqrt{-g}g^{tt}g^{uu} \phi'^2}{q^4(u)}.
\ee
From (\ref{SA}), we know the regularity at the horizon corresponds to
\be
J^x_k(1)=-i\omega {K(u)}\sqrt{\frac{g_{uu}}{-g_{tt}}}A_x(1).
\ee
The DC conductivity can be calculated by using the  following relation \cite{myers}
\be
\sigma_{DC}=-K(u)\sqrt{\frac{g_{uu}}{-g_{tt}}}
~~\bigg|_{u=1}\frac{A_x(1)A_x(1)}{A_x(u_c)A_x(u_c)} ,\label{DCexpress}\
\ee
where $A_x(u)$ is the solution of equation (\ref{EOMAxD}).
We can solve $A_x(u)$ by imposing  boundary condition at $u=0$
and set $\omega$ to zero ,  which leads to
\be
A_x(u)=A_x(0)\frac{(r_0^{2}+Q^2)^2}{(r_0^{2}+Q^2 u_c^2)}.  \label{solvefordc}\
\ee
Inserting (\ref{solvefordc}) into (\ref{DCexpress}) , finally we can
obtain the DC conductivity at the cutoff surface  (See Figure 1)
\be \label{dc}
\sigma_{DC}(u_c)=\frac{r_{0}}{2\kappa^2 L}(1+\frac{Q^2}{r_0^{2}})^{-1}(1+\frac{Q^2 u_c^2}{r_0^{2}})^2.
\ee
  At the horizon $u_c=1$, the above formula becomes
\be \sigma(u=1)=\frac{r_{0}}{2\kappa^2 L}(1+\frac{Q^2}{r_0^{2}}), \label{dch}
\ee
which agrees with (\ref{SA}).
In the boundary $u_c\rightarrow 0$, the DC
conductivity is reduced to
\be
\sigma(u=0)=\frac{r_{0}}{2\kappa^2 L}(1+\frac{Q^2}{r_0^{2}})^{-1}.
\ee
Without the chemical potential, (\ref{dc})  reduces to the formula for DC
conductivity given in \cite{Hong.Membrane}.

\begin{figure}[hbtp]
~~~~~~~~~~~~~~~~~~~~~~~~~\includegraphics*[bb=0 0 350 200,width=0.65\columnwidth]{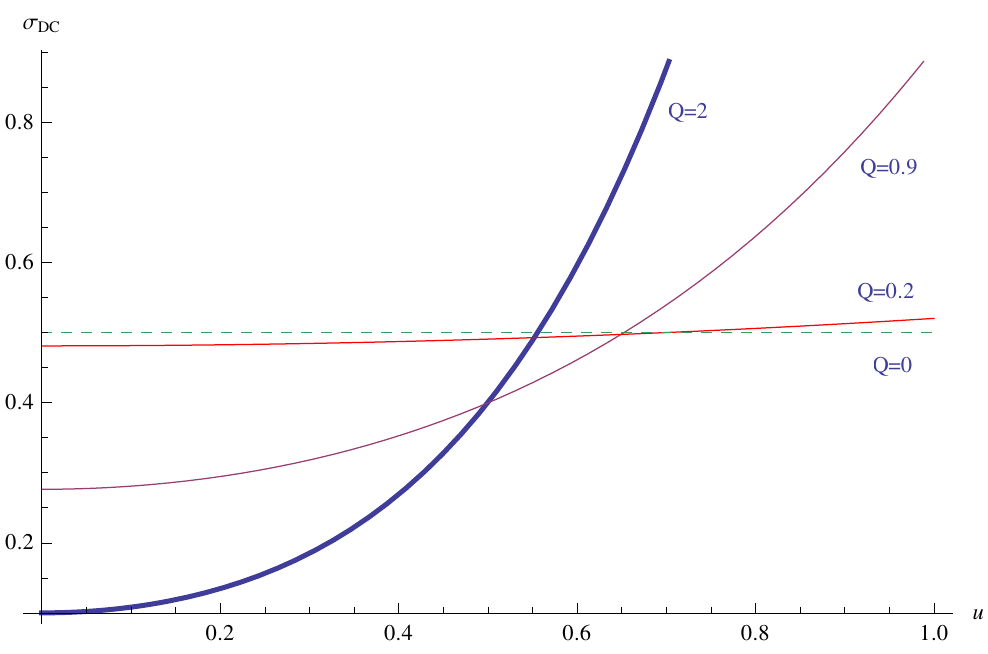}
\caption{ The DC conductivity  as a function of the radial coordinate $u$ with different charges, where we set $L=r_0=1=\kappa^2=1$. The dashed line corresponds to $Q=0$.
  \label{fig1}}
\end{figure}
 We can see from Fig. \ref{fig1} that the DC conductivity on the boundary depends on the charge $Q$. The slope of lines is proportional to the charge $Q$.

As a side note, we will check whether the transport coefficients calculated  satisfies the Einstein relation  $\sigma_{DC} / \chi \mid_{\omega=0}=D_c$. By using the relation which the value of susceptibility is defined by (\ref{sus}) , we find the expression for $\sigma_{DC} / \chi$ can be written as
\be
\frac{\sigma_{DC}(0)}{\chi}=\frac{r_0^3}{2(Q^2+r_0^2)(3Q^2+r_0^2)}.
\ee
We will see later that the right hand of equation is not the diffusion constant and thus the Einstein relation is not satisfied. In  the absence of the chemical potential $\mu=0$ (i.e. $Q=0$), the above equation becomes $\sigma_{DC} / \chi \mid_{\omega=0}=D_c$ with
$D_c=1/2\pi T$.

As a by-product,  we calculate the thermal conductivity by using the relation
\be
\kappa_T=(\frac{\epsilon+P}{T_H \rho})^2 T_H\sigma_{\rm DC}(0)=\frac{\pi r_0^4}{\kappa^2L^3Q^2}(1+\frac{Q^2}{r_0{^2}}).
\ee
The ratio $\frac{\kappa_T \mu^2}{\eta T_H}$ can be computed as
\be
\frac{\kappa_T \mu^2}{\eta T_H}=4\pi^2.
\ee

\subsubsection{Diffusion Coefficient}
Now we are going to calculate the diffusion coefficient. Let us evaluate the ``conductivity'' introduced by the metric
perturbation $h^x_z$ at any  momentum. The conductivity is defined as
\be
\sigma_h:=\frac{j^z}{f_{zt}}. \label{bb}
\ee In the absence of the  momentum,  the decoupled flow equation for $\sigma_h$ yields
\be
\frac{\partial_{u_c}\sigma_h}{-i\omega}+\sigma^2_h\frac{g_{uu}g_{zz}}{G}+G g^{tt}g^{zz}=0.
\ee
 Again, the regularity  condition at the event horizon gives
\be
\sigma_h(1)=\frac{r_{0}^3}{2\kappa^2 L^3}(1+\frac{Q^2}{r_0^{2}}). \label{boundry}\
\ee
Actually, the value of this conductivity is equivalent to shear viscosity.
Noting that from (\ref{bb}) and (\ref{boundry}), we can recast the
the conductivity at the horizon as
\be\label{bbb}
j^z(1)=\sigma_h(1)f_{zt}(1).
\ee

For the fields and equations of motion, we treat the vector modes in a long wave-length
expansion. We will find that the
diffusion constant  depends on $u_c$, charge and dilaton field. The equation of motion of $\sigma_h$ is coupled
with other modes when momentum $k_z$ is non-zero. To proceed, we take the scaling limit
for temporal and spatial derivatives as
\be
\partial_t \sim \epsilon^2,~~~~ \partial_z \sim \epsilon, ~~~ \\
{f}_{zt}\sim \epsilon^3\bigg({f}^{(0)}_{zt}+\epsilon
{f}^{(1)}_{zt}+... \bigg).
\ee
The in-falling boundary condition at the horizon implies $j^z$ is linearly related to
${f}_{zt}$. So, we have
\be
j^z\sim \epsilon^3\bigg(j^{z(0)}+\epsilon j^{z(1)}+...\bigg).
\ee
Through the charge conservation equation (\ref{heom0}), we have
\be
\tilde{j}^t \sim \epsilon^2 \bigg(\tilde{j}^{t(0)}+\epsilon \tilde{j}^{t(1)}+...\bigg)~~\\~~
j^t\sim \epsilon^2 \bigg({j}^{t(0)}+\epsilon j^{t(1)}+...\bigg).
\ee
To zeroth order, (\ref{heom1}) and (\ref{heom2}) can be reduced to
\be\label{lowestheom}
\partial_{u}\tilde{j}^{t(0)}=0,~~~\partial_{u}j^{z}=0.
\ee
The first equation of (\ref{lowestheom}) figures that
\be\label{tjt}
\tilde{j}^{t(0)}={j}^{t(0)}-\frac{1}{q^2}\sqrt{-g} g^{uu}g^{tt}\p_u A_t A^{(0)}_x=C_0,
\ee
where $C_0$ is a constant. The
equation (\ref{heom0}) implies
\be
j^{z(0)}=\tilde{j}^{t(0)}\frac{\omega}{k}.
\ee
The zeroth order of the Bianchi identity becomes
\be\label{bianchi0}
\partial_u {f}^{(0)}_{zt}+\frac{g_{uu}g_{tt}}{G}\partial_z {j}^{t(0)}=0.
\ee
Integrating the above equation from $u$ to the horizon $u=1$, we obtain
\be\label{tfzt}
 {f}^{(0)}_{zt}(u)= {f}^{(0)}_{zt}(u)\bigg|_{u=1}+2\kappa^2\int^1_{u}du\frac{\partial_z {j}^{t(0)}}{g^{uu}g^{tt}\sqrt{-g} g_{xx}}.
\ee
For the gauge perturbation $A^{(0)}_x$, the Maxwell equation in the scaling limit becomes
\be\label{Ax}
\partial_u\bigg(\frac{1}{q^2(u)}\sqrt{-g}g^{uu}g^{xx}\partial_u A^{(0)}_x\bigg)=\frac{1}{q^2(u)}\sqrt{-g}  g^{uu}g^{tt}\p_u A_t{{h}^x_t}'^{(0)}.
\ee
Following \cite{Andy.RG}, imposing the boundary condition
$A^{(0)}_x(u_c)=0$, and solving equation (\ref{tjt}) and (\ref{Ax}), we obtain
 \be
A^{(0)}_x(u)=\frac{Qr_0(2Q^2u_c^2+\frac{r_0^2}{u^2}+r_0^2u_c^2)}{4\sqrt{2}(Q^2+r_0^2)^2}(\frac{u^2}{u_c^2}-1),
\ee
and substitute it into (\ref{tjt}).
Consequently,  we have
\be\label{lf1}
f_{zt}(u)={f}^{(0)}(1)+\frac{i k
(1-u_c^2)(2Q^2u_c^2+(1+u_c^2)r_0^2)r_0L^2}{4(Q^2u_c^2+r_0^2) (Q^2+r_0^2)}.
\ee
Following the sliding membrane paradigm \cite{Hong.Membrane},  we defined the momentum conductivity by
the current and electric fields as
\be
\sigma_h (u_c)=\frac{j^{z(0)}(u_c)}{f^{(0)}_{zt}(u_c)}=\frac{\omega}{k}\frac{\tilde{j}^{t(0)}(u_c)}{f^{(0)}_{zt}(u_c)}.
\ee
By  using (\ref{lf1}) and the boundary condition given in (\ref{bbb}), we find the expression of the conductivity $\sigma_h$ satisfying
\be\label{cond}
\frac{1}{\sigma_h(u_c)}=\frac{1}{\sigma_h(1)}-\frac{k^2}{i\omega}\frac{{D}(u_c)}{\sigma_h(1)},
\ee
where the diffusion constant is given by
\be
D(u_c)=\frac{r_0L^2(2Q^2u_c^2+(1+u_c^2)r_0^2)}{4(Q^2u_c^2+r_0^2) (Q^2+r_0^2)}(1-u_c^2).
\ee
Note that the dimensional $D(u_c)$ can be set to any value by a coordinate transformation. So it is necessary to obtain a dimensionless diffusion constant in the following sections.

We can define the proper frequency $\omega_c$ and  the
proper momentum $k_c$  on the hypersurface
$u=u_c$
\be
\omega_c\equiv \frac{\omega}{\sqrt{-g_{tt}}},~~~~~~~k_c\equiv
\frac{k}{\sqrt{g_{ii}}},
\ee
which is conjugate to  the  proper time and
the proper  distance, respectively. By using the Tolman relation, we can obtain the  Hawking temperature at the cut-off surface $u=u_c$ which is
$T_c(u_c)=\frac{T}{\sqrt{-g_{tt}}}$.

We define a dimensionless diffusion constant $\bar{D}(u_c)$ which is coordinate-invariant as
\be
\bar{D}(u_c)=D(u_c)T_c \frac{g_{zz}}{\sqrt{-g_{tt}}}.
\ee
Thus, in terms of the normalized momentum and the diffusion
constant, the conductivity can be expressed as
\be\label{cond1}
\frac{\sigma_h(1)}{\sigma_h(u_c)}=1-\frac{k^2_c}{i\omega_c}\frac{\bar{D}(u_c)}{T_c},
\ee
where the dimensionless diffusion coefficient is given by  (See Figure 2)
\be\label{bd1}
\bar{D}(u_c)=\frac{1}{4\pi}\frac{(Q^2u_c^2+r_0^2)}{(Q^2+r_0^2)}.
\ee

\vspace{.5cm}
\begin{figure}[hbtpq]
~~~~~~~~~~~~~~~~~~~~~~~~~\includegraphics*[bb=0 0 300 200,width=0.65\columnwidth]{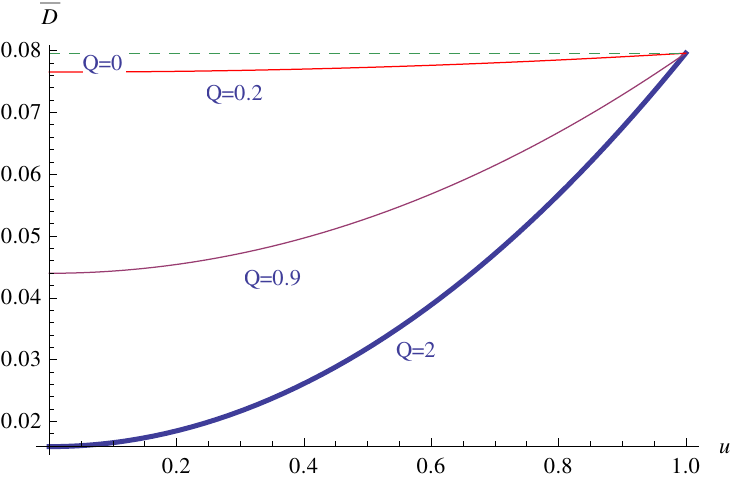}
\caption{The dimensionless diffusion coefficient as a function of radial coordinate $u$ with $L=r_0=1=\kappa^2=1$ .
  \label{fig2}}
\end{figure}

Fig.\ref{fig2} shows the dimensionless diffusion coefficient of charged dilatonic black hole runs for different charges on cut-off surface.  The dash line shows the dimensionless diffusion coefficient is a constant $\frac{1}{4\pi}$ when $Q=0$. On the boundary, different charge gives different diffusion constant while at the horizon $u=1$, all these lines end at the same value $\bar{D}=1/4 \pi$.  Note that the slope of these lines is also proportional to the charge $Q$.

\subsection{Transport Coefficients From Brown-York Stress Tensor}
In this section, we will provide a consistent check by using the black hole thermodynamics. We will verify that by using the formula
\bea \label{itis}
&&\sigma_{DC}{(u_c)}=\frac{\Omega(1)}{\digamma^2(u_c)},
\eea
one can easily obtain  the DC conductivity. For a $d+2$-dimensional charged black holes, the dimensionless diffusion coefficient can be written in a simple form
\bea
&&\bar D(u_c)=\frac{1}{4\pi}\frac{\Xi(1)}{\digamma(u_c)},
\eea
where
\bea
&&\Xi(1)=\frac{-g^3_{xx}g'_{tt}}{\sqrt{-g}}\bigg|_{u=1} ,\nonumber \\
&&\Omega(1)=\bigg(\frac{g^{(5d-2)/4}_{xx}g'_{tt}}{q(u)\sqrt{-g}}\bigg)^2\bigg|_{u=1},\nonumber \\
&&\digamma(u_c)=\sqrt{-g}g^{uu}\bigg(\ln {\frac{-g_{tt}}{g_{xx}}}\bigg)'\bigg|_{u=u_c}.\nonumber
\eea
On an arbitrary cutoff surface $u=u_{c}$ outside the horizon, there
is  a thermodynamic description of the
fluid dual to the background configuration (\ref{Onechargeexpress}). The Brown-York tensor is defined as \cite{B-Y}
\begin{equation}
T_{ij}=\frac{1}{8\pi G_5}(\Theta \gamma_{ij}-\Theta_{ij}-\mathcal{C}\gamma_{ij})\label{BYD},
\end{equation}
with the extrinsic curvature
\begin{equation}
\Theta_{ij}=-\frac{1}{2}(\nabla_i n_j+\nabla_j n_i), ~~~i,j=t,x_i\label{excurv}.
\end{equation}
The induced metric $\gamma_{\mu\nu}$ can be expressed by bulk metric $g_{\mu\nu}$ and the normal vector in $u$ direction, so
 \begin{equation}
\gamma_{\mu\nu}=g_{\mu\nu}-n_\mu n_\nu,
\end{equation}  where
 $n^u=\frac{u^2\sqrt{h}}{r_0e^B}$ is a unit normal vector, and  the trace of extrinsic curvature is given by
$\Theta\equiv \gamma^{ij}\Theta_{ij}$. One may note that the indices $\mu, \nu $ of the bulk and the indices $i,j$, which  live  on the cutoff surface. By calculating with the metric and induced metric, we obtain
\begin{equation}
    \Theta_{tt}=-\frac{\sqrt{h}u^2}{r_0e^B}(\frac{h'}{2h}+A')\gamma_{tt}, \qquad \Theta_{xx}=-\frac{\sqrt{h}u^2A'}{r_0e^B}\gamma_{xx}, \qquad \Theta=-\frac{\sqrt{h}u^2}{r_0e^B}(\frac{h'}{2h}+4A').
\end{equation}
The stress tensor at an arbitrary cut-off surface can be written as
\begin{equation}
T_{ij}d{x}^{i}d{x}^{j} = \frac{1}{8\pi G_5}\bigg[\frac{\sqrt{h}h u^2e^{2A}}{r_0e^B}3A' dt^{2}-\frac{\sqrt{h}u^2e^{2A}}{r_0e^B}(3A'+\frac{h'}{2h}) d{x}_{i}d{x}^{i}-\mathcal{C}ds_{c}^{2}\bigg].  \label{BY metric}
\end{equation}
 On the other hand, the stress-energy tensor of a relativistic fluid in
equilibrium is
\begin{equation}
T_{ij}=(\varepsilon+P)u_{i}u_{j}+P \gamma_{ij},\label{BY tensor}
\end{equation}
where $\varepsilon$ denotes the energy density, $P$ the pressure and
$u^{i}=\{\frac{e^{-A}}{\sqrt{h}},0,0,0\}$ the normalized fluid
four-velocity. For simplicity, we can redefine the energy density and pressure
\begin{equation}
P\rightarrow P-\frac{\mathcal{C}}{8\pi
G_5},\qquad \epsilon\rightarrow \varepsilon+\frac{\mathcal{C}}{8\pi G_5},
\end{equation}
which leaves the combination $\varepsilon+P$ invariant. In this sense, we can omit the constant $\mathcal{C}$ in the Brown-York tensor\footnote{Note that $\mathcal{C}$  is  the counterterm of the action and essential for the regularity of $T_{ij}$
as $u_{c}\to 0 $.}. Comparing (\ref{BY metric}) with (\ref{BY tensor}), we find
\begin{equation}
\varepsilon+P=\frac{1}{16\pi G_5}n^u(\frac{\gamma'_{tt}}{\gamma_{tt}}-\frac{\gamma'_{xx}}{\gamma_{xx}})=-\frac{u_c^2h'}{2\kappa^2\sqrt{h}r_0e^B}\bigg|_{u=u_c}.\label{energeprssurec}
\end{equation}
Notice that the entropy density is given by
\begin{equation}
s_c=\frac{A_{H}}{4G_5 V_3}=\frac{2\pi r_0\sqrt{m}}{\kappa^2L^2e^{3A}}\bigg|_{u=u_c} \label{entropyc}.
\end{equation}
 The local Hawking temperature is given by
\begin{equation}\label{tempc}
T_c=\frac{T_{H}}{\sqrt{-g_{tt}}}=\frac{r_0}{\pi L^2\sqrt{h}e^A}\bigg|_{u=u_c} ~~~  , ~~~T_H=\frac{1}{4\pi}\frac{-g'_{tt}}{\sqrt{-g_{tt}g_{uu}}}\bigg|_{u=1}.
\end{equation}
We can express thermodynamic variables in terms of metric components
\begin{equation}
\frac{s_cT_c}{\varepsilon+P}=-\frac{4r_0^3\sqrt{m}}{u_c^2L^4h'e^{4A-B}}\bigg|_{u=u_c},
\end{equation}
By detailed calculation, we can obtain the value
\begin{equation}
\frac{s_cT_c}{\varepsilon+P}=\frac{Q^2u_c^2+r_0^2}{Q^2+r_0^2}.
\end{equation}
Apparently, one can verify the following identity
\be \label{whata}
\frac{s_cT_c}{\varepsilon+P}=\frac{\Xi(1)}{\digamma(u_c)},
\ee by introducing the quantities
\be
\Xi(1)=\frac{-g^3_{xx}g'_{tt}}{\sqrt{-g}}\bigg|_{u=1} ,\nonumber
\ee
\be
\Omega(1)=\bigg(\frac{g^{(5d-2)/4}_{xx}g'_{tt}}{q(u)\sqrt{-g}}\bigg)^2\bigg|_{u=1}, \nonumber
\ee
\be
\digamma(u_c)=\sqrt{-g}g^{uu}\bigg(\ln {\frac{-g_{tt}}{g_{xx}}}\bigg)'\bigg|_{u=u_c}  \nonumber
\ee
Finally, we verify that
\begin{equation}
\sigma_{DC}{(u_c)}=\frac{\Omega(1)}{\digamma^2(u_c)}.
\end{equation}
In concrete, it can be written as
\begin{equation}\label{condh}
\sigma_{DC}{(u_c)}=\frac{r_{0}}{2\kappa^2 L}\bigg(1+\frac{Q^2}{r_0^{2}}\bigg)^{-1}\bigg(1+\frac{Q^2 u_c^2}{r_0^{2}}\bigg)^2.
\end{equation}
The above result agrees with (\ref{dc}).

Moreover, we can verify the diffusion constant by using
\be\label{df}
\bar{D}(u_c)=\frac{\eta}{\varepsilon+P}T_c=\frac{1}{4\pi}\frac{s_cT_c}{\varepsilon+P}, ~~~~
\ee where $\eta(u_c)$ is to be obtained from (\ref{cutoffeta1}). Note that throughout the paper, we consider the case
in which the KSS bound is satisfied.
In terms of the metric component, the diffusion coefficient can be expressed as
\be
\bar{D}(u_c)=\frac{\Xi(1)}{4\pi\digamma(u_c)}
\ee
For our case, the diffusion coefficient is given by
\be
\bar{D}(u_c)=\frac{1}{4\pi}\frac{(Q^2u_c^2+r_0^2)}{(Q^2+r_0^2)}.
\ee  This is  consistent with (\ref{bd1}). As $u_c\rightarrow 1$, the diffusion constant becomes $\bar{D}(u_c)=\frac{1}{4\pi}$.

\section{R-charged Black Holes}

The hydrodynamics of R-charged black holes was studied by several authors \cite{myers,Rcharged,gem,erdmenger,jain}.
The retarded Green functions were evaluated  on the  boundary and the  boundary theory transport coefficients were computed up to first and second orders.
In order to prove that the formula conjectured for transport coefficients has their universal applications, we will compute the RG flow of the single-charged black holes in the following.

We know the effective Lagrangian of a single charged black hole can be written as \cite{STU}
\begin{equation}
{{\cal L}\over \sqrt{-g}} =
 R + {2\over L^2} {\cal V} -\frac{L^2}{8} H^{4/3} F^2
-\frac{1}{3}
H^{-2} g^{\mu\nu} \partial_\mu H\, \partial_\nu H \,,
\label{SClagrangian_1}
\end{equation}
where the potential for the scalar field $H$ and ${\cal V}$ is,
\begin{equation}
{\cal V} = 2 H^{2/3}+4 H^{-1/3}\,.
\label{SCpot}
\end{equation}
The single-charged metric and gauge fields are given by
\begin{eqnarray}
ds_5^{2} &=& \frac{\left(\pi  {\mathcal T}_0 L \right)^{2} }{u}H^{1/3}
\left( \frac{-f}{H} dt^{2}+dx^{2}+dy^{2}+dz^{2}\right)+\frac{L^{2}}{4fu^{2}} H^{1/3}du^{2}, \label{SCmetrics}
\nonumber\\
A_{\mu } &=& \pi  {\mathcal T}_0\sqrt{2\kappa_1 \left( 1+\kappa_1 \right) }\frac{u}{H}\left(dt \right) _{\mu },
\end{eqnarray}
where
\begin{equation}
H = 1 + \kappa_1 u,~~~
f = \left(1-u \right)\left( 1 + \left(1 +\kappa_1  \right)u  \right),~~
T = \frac{1 + \frac{\kappa_1 }{2}}{\sqrt{1+\kappa_1 }}\mathcal T_0,~~ \mathcal T_0=\frac{r_{0}}{\pi L^2},
\end{equation}
with $\kappa_1$ denoting the R-charge while $\mathcal T_0$ denotes the Hawking temperature of neutral black hole.
The energy density, pressure and entropy density are given by
\be
\epsilon=\frac{3\pi^2 N^2 T^4_0}{8}(1+\kappa_1),~~~P=\frac{\epsilon}{3},~~~s=\frac{\pi^2 N^2 T^3_0}{2}\sqrt{1+\kappa_1}.
\ee
The charge density and conjugated chemical potential are
\be
\rho=\frac{\pi N^2 T^3_0}{8}\sqrt{2\kappa}\sqrt{1+\kappa_1}, ~~~\mu=\frac{\pi T_0 \sqrt{2\kappa}}{\sqrt{1+\kappa_1}}.
\ee
The Newton constant is given by $G_5=\frac{\pi L^3}{N^2}$.
The system of the  gauged supergravity equations of motion for the fields
$g_{\mu\nu}$, $A_{\mu}$, $H$ read
\begin{equation}
\begin{split}
\,& \Box H = H^{-1} g^{\mu\nu} \partial_\mu H \, \partial_\nu H
+\frac{L^2}{4} H^{7/3} F^2-\frac{3}{L^2} H^2 \frac{\partial \cal V}{\partial H}\,,\\
\,&\partial_\mu\left(\sqrt{-g} H^{4/3} F^{\mu\nu}\right) =0\,,\\
\,& R_{\mu\nu}=\frac{L^2}{4} H^{4/3} F_{\mu\gamma} F_{\nu}\ ^\gamma
+\frac{1}{3} H^{-2} \partial_\mu H \partial_\nu H-g_{\mu\nu}[
\frac{2}{3 L^2} {\cal V}+ \frac{L^2}{24}H^{4/3}F^2 ]\,.
\end{split}
\label{SCeom1}
\end{equation}
The vector type perturbation takes the form
\begin{equation}
h_{tx} =	g_{xx}\left( u\right) h^x_t\left( u\right) e^{-i\mathfrak{w} t + iKz},
 h_{zx} =	g_{xx}\left( u\right) h^x_z\left( u\right) e^{-i\mathfrak{w} t + iKz},
 a_{x} = \frac{\mu }{2}A_x\left( u\right) e^{-i\mathfrak{w} t + iKz},
\end{equation} where $\omega = \frac{\mathfrak{w} }{2\pi {\mathcal T}_0}$ and $q = \frac{K  }{2\pi {\mathcal T}_0}$ are dimensionless frequency and momentum.
We are interested in gravitational fluctuations of the shear type, where
the only nonzero components of $h_{\mu\nu}$ are $h_{t a}$, $h_{z a}$, $a=x,y$.
One can show that fluctuations of all other fields except
$A_a (r,t,z)$, $a=x,y$, can be consistently set to zero.
Introduce the new variables
\begin{equation}
h^x_t\equiv H_{ta} = g^{x x} h_{t a}\,,\;\;\;   h^x_z \equiv H_{za}
= g^{x x} h_{z a}\,,\;\;\;
A_x = {2 A_a\over \mu}.
\label{SCrescaling}
\end{equation}
The linearized equations derived from Eqs.~(\ref{SCeom1}) can be written as\cite{Rcharged}
\begin{subequations}
\begin{eqnarray}
&& {h^x_t}' + {q \,  f \over \omega \, H } \; {h^x_z}' +
{\kappa_1  u \over 2 \, H}\;  A_x =0\,,
\label{SCeqq1r}\\
&& {h^x_t}'' + {u  H' -  H\over u H}\; {h^x_t}' - { \omega \, q \over f u} \; {h^x_z} -
 { q^2\over f u} \; {h^x_t} + {\kappa_1 \, u\over 2\,  H}
 A_x ' =0\,,  \label{SCeqq2r}\\
&& {h^x_z}'' + {u f' - f \over u f}
\; {h^x_z}' +  { \omega^2 \, H\over  f^2 u}\; {h^x_z}
+  { q \, \omega \, H\over  f^2 u} \; h^x_t =0\,,  \label{SCeqq3r}\\
&&
\left( H f A_x' + 2(1+ \kappa_1)\, h^x_t \right)' -  { q^2 H\over u}\; A_x
+ { \omega^2 H^2\over  f u} \; A_x =0\,. \label{SCeqq4r}
\end{eqnarray}
\end{subequations}
The above four equations are not independent. Combining Eq.~(\ref{SCeqq1r})
with Eq.~(\ref{SCeqq2r}), one obtains Eq.~(\ref{SCeqq3r}). Thus it is sufficient to
consider
 Eqs.~(\ref{SCeqq1r}), (\ref{SCeqq2r}),
(\ref{SCeqq4r}).
In the zero momentum limit, Eq.(\ref{SCeqq1r}) becomes
\begin{equation}
    {h^x_t}'=-\frac{u\kappa_1}{2H}A_x,  \label{SCRdecoupt}
\end{equation}
Substituting (\ref{SCRdecoupt}) into (\ref{SCeqq4r}), we obtain
\begin{equation}
    A_x''+(\frac{f'}{f}+\frac{H'}{H})A_x'+\frac{\omega^2H}{uf^2}A_x-\frac{u\kappa_1(1+\kappa_1)}{fH^2}A_x=0.  \label{SCRdecoupmaxw}
\end{equation}
Now we introduce  $j^\mu$ and $f_{\mu\nu}$ to denote the current and strength for the vector modes
\bea
&&j^\mu = -{1\ov g_{\text{eff}}^2(u)}\sqrt{-g}f^{u\mu}\ ,\quad f_{\mu\nu} = \p_\mu h^x_\nu - \p_\nu h^x_\mu\ ,\\
&&G =\frac{\sqrt{-g}}{g^2_{\text{eff}}(u)}\ , \quad  \frac{1}{q^2(u)} = {L^2H^{4/3}\ov 4\kappa^2}\ , \quad {1\ov g^2_{\text{eff}}(u)}= {(\pi  {\mathcal T}_0L)^2H^{1/3}\ov 2\kappa^2u}\ .
\eea
By further define
\be
 j^\lambda=- \frac1{q^2(u)}\sqrt{-g} \frac{4\kappa_1 u^3}{L^4H^{5/3}} A_x ,
\ee
and
\be
    \tilde{j^t} =  j^t +j^\lambda,\nonumber \\
\ee
We can recast the equations of motion as
\bea
  &&\p_t\tilde{j^t} + \p_z j^z = 0,\\ \label{SCheom0}\
  &&\p_u\tilde{j^t} + Gg^{tt}g^{zz}\p_zf_{zt} = 0,\\ \label{SCheom1}\
  &&\p_uj^z - Gg^{tt}g^{zz}\p_tf_{zt} = 0. \label{SCheom2}\
\eea
The Bianchi identity holds as
\be \label{SCheom3}
-{g_{uu}g_{zz}\ov G}\p_tj^z + {g_{uu}g_{tt}\ov G}\p_zj^t + \p_uf_{zt} = 0\ .
\ee
Similarly  we can define $j^x$ and $\tilde j^x$ as
 \be
 \tilde{j}^x = j^x +j^\rho,\quad
j^x = -\frac1{q^2(u)}\sqrt{-g} F^u{^x} \ , \quad
   \ee
  where
\be
j^\rho=\frac1{q^2(u)}\sqrt{-g}  \frac{8u^3(1+\kappa_1)}{(\pi  {\mathcal T}_0)^2L^4H^{5/3}} { h^x_t}'.
\ee
 The equation of motion for $A_x$ can be written as
\be
\label{SCAeomx}
-\p_u \tilde{j^x} + \frac{\sqrt{-g}}{q^2(u)}g^{xx}(g^{tt}\p_t F_{tx} - g^{zz}\p_z F_{zx}) = 0.
\ee
One can see
  that $  A_x $   decouples from $h^x_t$ in the $q\to 0$  limit
\be
\label{SCEOMAxD}
\p_u(\frac1{q^2(u)}\sqrt{-g}g^{uu}g^{xx}\p_u  A_x)- \frac{\sqrt{-g}}{q^2(u)} \frac{8u^4\kappa_1(1+\kappa_1)}{(\pi  {\mathcal T}_0)^2L^4H^{8/3}} A_x -\frac{\sqrt{-g}}{q^2(u)}
g^{xx}g^{tt}\omega^2 A_x =0\ .
\ee

\subsection{DC Conductivity}
 By defining
\be
   \sigma_x = {j^x \ov i\omega A_x},  \label{SCsigmadef}\
\ee
we re-arrange  (\ref{SCEOMAxD}) as the flow equation for electric conductivity
\be\label{SCsigmaflow0}
{\p_u \sigma_x\ov i\omega} -
{q^2(u)\sigma_x^2\ov \sqrt{-g}g^{uu}g^{xx}} +\frac{\sqrt{-g}}{q^2(u)}\frac{8u^4\kappa_1(1+\kappa_1)}{\omega^2(\pi  {\mathcal T}_0)^2L^4H^{8/3}}- \frac1{q^2(u)}\sqrt{-g}g^{xx}g^{tt} = 0\ .
\ee
We can immediately write down the regularity condition at the horizon
\be\label{SCSA}
\sigma_x(u=1)= \frac{1}{q^2(u)
}\sqrt{\frac{-g}{g_{uu}g_{tt}}}g^{xx}\bigg|_{u=1}= \frac{r_0}{16\pi G_5 L}(1+\kappa_1)^{\frac{3}{2}}.
\ee
It is convenient to define the radial momentum as
\be
J^x_k=\frac{1}{\kappa^2}K(u)A_x'(u,k).
\ee
The equation of motion for $A_x$ then takes the form \cite{ge3}
\be
\partial_uJ^x_k=\frac{1}{\kappa^2}L(u)A_x(u,k),
\ee
where
\be
K(u)=-\frac{\sqrt{-g}}{q^2(u)}g^{xx}g^{uu},~~~L(u)=\omega^2\frac{\sqrt{-g}}{q^2(u)}g^{xx}g^{tt}-\frac{\sqrt{-g}}{q^2(u)}\frac{8u^4\kappa_1(1+\kappa_1)}{(\pi  {\mathcal T}_0)^2L^4H^{8/3}}.
\ee
The regularity at the horizon $u=1$ corresponds to
\be
J^x_k(1)=-i\omega {K(u)}\sqrt{\frac{g_{uu}}{-g_{tt}}}A_x(1).
\ee
According to the following relation \cite{ge3}
\be
\sigma_{DC}=-K(u)\sqrt{\frac{g_{uu}}{-g_{tt}}}
~~\bigg|_{u=1}\frac{A_x(1)A_x(1)}{A_x(u_c)A_x(u_c)}, \label{SCDCexpress}\
\ee
where $A_x(u)$ is the solution of equation (\ref{SCeqq4r})
at zero momentum.
We can solve $A_x(u)$ by imposing  some regularity  condition at the horizon
and setting $\omega$ to zero ,  which lead to
\be
A_x(u)=A_x(0)\frac{2+u_c\kappa_1}{2(1+u_c\kappa_1)}.  \label{SCsolvefordc}\
\ee
Finally we can
obtain the DC conductivity at the cutoff surface (See Figure 3)
\be \label{SCdc}
\sigma_{DC}(u_c)=\frac{r_0}{16\pi G_5 L}(1+\kappa_1)^{\frac{3}{2}}\bigg(\frac{2+\kappa_1}{2+u_c\kappa_1}\bigg)^2\bigg(\frac{1+u_c\kappa_1}{1+\kappa_1}\bigg)^2.
\ee
  At the horizon $u_c=1$, the above equation becomes
\be \sigma_{\rm DC}=\frac{1}{16\pi G_5}\frac{r_0}{L}(1+\kappa_1)^{3/2}, \label{SCdch}
\ee
which is consistent with (\ref{SCSA}).
As one goes to the boundary $u_c\rightarrow 0$, the DC
conductivity is reduced to
\be
\sigma_{\rm DC}=\frac{1}{16\pi G_5}\frac{r_0}{L}(1+\kappa_1)^{-1/2}(1+\frac{\kappa_1}{2})^{2},
\ee
in good agreement with \cite{Rcharged}.

\vspace{.1cm}
\begin{figure}[hbtp]
~~~~~~~~~~~~~~~~~~~~~~~~~\includegraphics*[bb=0 0 350 200,width=0.65\columnwidth]{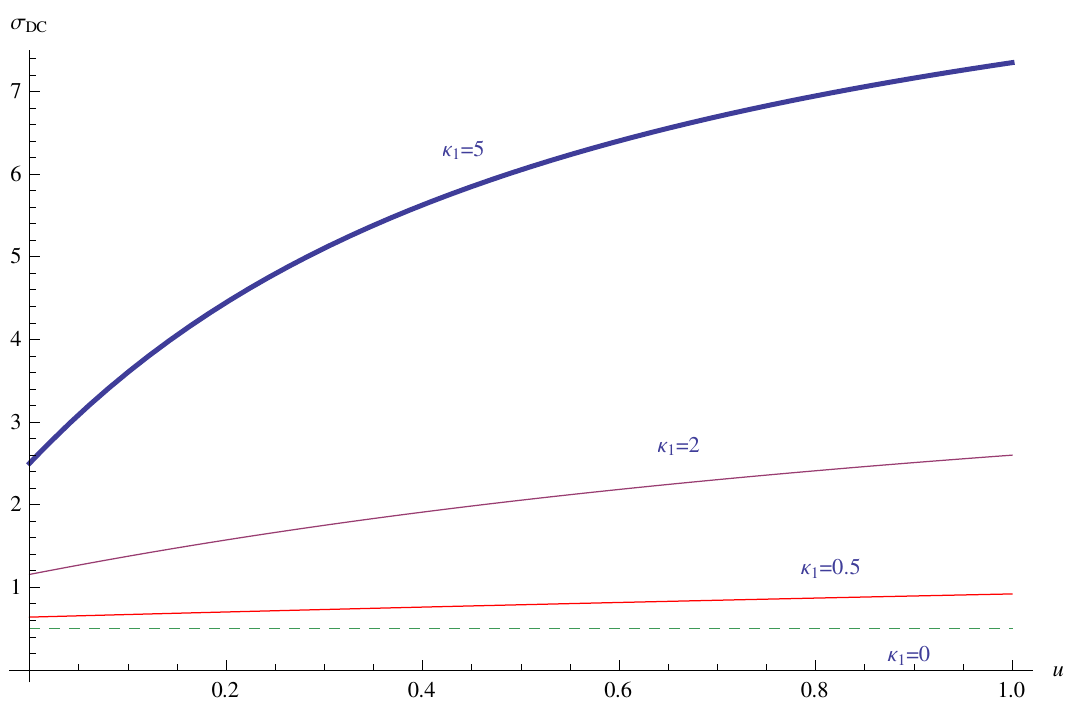}
\caption{ The DC charge conductivity as a function of the radial coordinate $u$. The dashed line represents the zero charge case
  \label{fig3}}
\end{figure}

Fig.\ref{fig3} shows the DC conductivity of single charged black hole runs  on a cut-off surface. We can see that for different charges, the DC conductivity on the boundary and at the horizon is different.

\subsection{Shear Viscosity}
We can write the same equation of motion according to minimally coupled massless scalar as:
\be
\partial_\mu (\sqrt{-g} \partial^\mu h^y_x)=0\ .\label{SCtensormode}\
\ee
The flow equation for shear viscosity is the same as in the previous section
\be
\label{SCetaflow}
{\p_{u_c}\eta(u_c, \omega)}=i\omega \left({2\kappa^2 \eta^2(u_c, \omega)\ov \sqrt{-g}g^{uu}}-\frac{\sqrt{-g}g^{tt}}{2\kappa^2}\right)\ .
\ee
On the horizon, the regularity gives
\be
\label{SCfloweta}
\eta(r_0)=\frac{{r_0}^3}{2\kappa^2 L^3} (1+\kappa_1)^{1/2}.
\ee
The entropy density is
\be
s={2  \pi r_0^3  \over \kappa^2 L^3}(1+\kappa_1)^{1/2}.
\ee
So the shear viscosity to entropy density ratio $\frac{\eta}{s}=\frac{1}{4\pi}$.
 Following \cite{Andy.RG}, we assume that this ratio does not change  with cut-off surface. Under this assumption, we obtain  the shear viscosity of any cut-off surface as
\begin{equation}
 \eta_c(u_c) =\frac{u_c^{3/2}}{2\kappa^2}\frac{(1+\kappa_1)^{1/2}}{(1+u_c\kappa_1)^{1/2}}. \label{SCcutoffeta}
\end{equation}
 On any cutoff surface, the entropy density can be expressed as
\begin{equation}
  s_c(u_c)=\frac{u_c^{3/2}}{4G_5}\frac{(1+\kappa_1)^{1/2}}{(1+u_c\kappa_1)^{1/2}}.
\end{equation}
\subsection{Diffusion Coefficient}
In order to obtain the diffusion coefficient, we need to evaluate the ``conductivity'' introduced by the metric
perturbation $h^x_z$ at any  momentum. The conductivity in this case
can be defined as
\be
\sigma_h:=\frac{j^z}{f_{zt}}. \label{SCbb}
\ee In the zero momentum
limit,  the decoupled flow equation for $\sigma_h$ is given by
\be
\frac{\partial_{u_c}\sigma_h}{-i\omega}+\sigma^2_h\frac{g_{uu}g_{zz}}{G}+G g^{tt}g^{zz}=0.
\ee
The regularity  condition at the event horizon gives
\be\label{SCboundry}
\sigma_h(1)=\frac{r_{0}^3}{2\kappa^2 L^3}(1+\kappa_1)^{1/2}.
\ee
It is worth noting that (\ref{SCbb}) and (\ref{SCboundry})
indicates that the conductivity at the horizon obeys
\be\label{SCbbb}
j^z(1)=\sigma_h(1)f_{zt}(1).
\ee
We take  the scaling limit
for temporal and spatial derivatives as
\be
\partial_t \sim \epsilon^2,~~~~ \partial_z \sim \epsilon, ~~~ \\
{f}_{zt}\sim \epsilon^3\bigg({f}^{(0)}_{zt}+\epsilon
{f}^{(1)}_{zt}+... \bigg).
\ee
With regard to the  lowest order, we have $\partial_{u}\tilde{j}^{t(0)}=0,~\partial_{u}j^{z}=0$.
The equation of $\tilde{j}^{t(0)}$ indicates that
\be\label{SCtjt}
\tilde{j}^{t(0)}={j}^{t(0)}- \frac1{q^2(u)}\sqrt{-g} \frac{4\kappa_1 u^3}{L^4H^{5/3}} A_x=C_0.
\ee
Note that $C_0$ is a constant. For the gauge perturbation
$A^{(0)}_x$, the Maxwell equation becomes
\be\label{SCAx}
\partial_u\bigg(\frac{1}{q^2(u)}\sqrt{-g}g^{uu}g^{xx}\partial_u A^{(0)}_x\bigg)=\frac1{q^2(u)}\sqrt{-g}  \frac{8u^3(1+\kappa_1)}{(\pi  {\mathcal T}_0)^2L^4H^{5/3}} { h^x_t}'^{(0)}.
\ee
The zeroth order Bianchi identity yields
\be\label{fzt}
{f}^{(0)}_{zt}(u)= {f}^{(0)}_{zt}(u)\bigg|_{u=1}+2\kappa^2\int^1_{u}du\frac{\partial_z {j}^{t(0)}}{g^{uu}g^{tt}\sqrt{-g} g_{xx}}.
\ee After imposing the boundary condition
$A^{(0)}_x(u_c)=0$ , we solve (\ref{SCAx}) and obtain the solution for $A^(0)_x$
 \be
A^{(0)}_x(u)=(1-\frac{u}{u_c})\bigg[\frac{3}{u}+8(1+\kappa_1)u_c-\frac{(2+\kappa_1)^2(4u_c-\kappa_1^2u_c^2u)}{(2+u\kappa_1)^2(2+u_c\kappa_1)^2}\bigg].
\ee
Then substituting it into (\ref{fzt}), we have
\be\label{SClf1}
f_{zt}(u)={f}^{(0)}(1)+i k
\frac{L^2(2+\kappa_1)(1+u\kappa_1)(1+u+u\kappa_1)}{2r_0(1+\kappa_1)^{3/2}(2+u\kappa_1)^2}(1-u_c).
\ee
Following the sliding membrane paradigm \cite{Hong.RG},  we define  the momentum conductivity by
the current and electric fields as
\be
\sigma_h (u_c):=\frac{j^{z(0)}(u_c)}{f^{(0)}_{zt}(u_c)}=\frac{\omega}{k}\frac{\tilde{j}^{t(0)}(u_c)}{f^{(0)}_{zt}(u_c)}.
\ee
By further using the boundary condition given in (\ref{SCbbb}), we find the expression for the conductivity $\sigma_h$ from (\ref{SClf1})
\be\label{SCcond}
\frac{1}{\sigma_h(u_c)}=\frac{1}{\sigma_h(1)}-\frac{k^2}{i\omega}\frac{{D}(u_c)}{\sigma_h(1)},
\ee
where the dimensional diffusion is
\be
D(u_c)=\frac{L^2(2+\kappa_1)(1+u\kappa_1)(1+u+u\kappa_1)}{2r_0(1+\kappa_1)^{3/2}(2+u\kappa_1)^2}(1-u_c).
\ee
The dimensionless diffusion constant $\bar{D}(u_c)$ can be defined as
\be
\bar{D}(u_c)=D(u_c)T_c \frac{g_{zz}}{\sqrt{-g_{tt}}},
\ee
where $
T_c=\frac{T}{\sqrt{-g_{tt}}}.
$
Thus the conductivity can be written in terms of the normalized momentum and the diffusion
constant as follows
\be\label{SCcond1}
\frac{\sigma_h(1)}{\sigma_h(u_c)}=1-\frac{k^2_c}{i\omega_c}\frac{\bar{D}(u_c)}{T_c},
\ee
where the dimensionless diffusion coefficient is given by (See Figure 4)
\be\label{SCbd1}
\bar{D}(u_c)=\frac{(2+\kappa_1)}{(2+u_c\kappa_1)} \frac{(1+u_c\kappa_1)}{4\pi(1+\kappa_1)}.
\ee
When $u_c\rightarrow 0$, the diffusion coefficient reduce to
\be
\bar{D}(0)=\frac{2+\kappa_1}{8\pi(1+\kappa_1)},
\ee
which exactly agrees with \cite{Rcharged}. As $u_c \rightarrow 1$, the diffusion constant becomes
\be
\bar{D}(1)=\frac{1}{4\pi}.
\ee

\vspace{.5cm}
\begin{figure}[hbtp]
~~~~~~~~~~~~~~~~~~~~~~~~~\includegraphics*[bb=0 0 350 200,width=0.65\columnwidth]{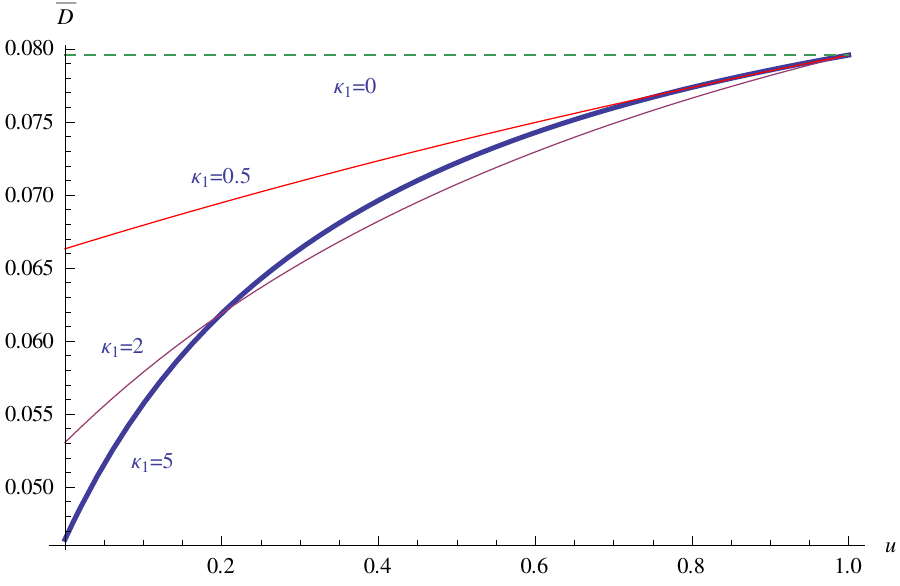}
\caption{ The dimensionless diffusion coefficient of single charge runs on the cutoff surface with different charges. The dashed line corresponds to the chargeless case.
  \label{fig4}}
\end{figure}

From Fig.\ref{fig4} we can see that while  the diffusion coefficientd is  different at the boundry, they all approach to the same value at $u=1$. Also different charges can  affect the diffusion coefficient seriously. When the charge becomes more larger, the ratio of diffusion coefficient to the cutoff radius $u$ become bigger.

\subsection{Transport Coefficients From Brown-York Stress Tensor}
In the following, we present a consistent check by using the Brown-York tensor.
The Brown-York tensor is defined as
\begin{equation}
T_{ij}=\frac{1}{8\pi G_5}(\Theta \gamma_{ij}-\Theta_{ij}-\mathcal{C}\gamma_{ij})\label{SCRBYD}.
\end{equation}
  The induced metric $\gamma_{\mu\nu}$ can be expressed by bulk metric $g_{\mu\nu}$ and the normal vector in $u$ direction. The norm vector is
 $n^u=\frac{2\sqrt{f}u}{LH^{1/6}}$  and  the trace of extrinsic curvature is given by
$\Theta\equiv \gamma^{ij}\Theta_{ij}$.  By calculating with the metric and induced metric , we obtain
\bea
  &&\Theta_{tt}=\frac{\sqrt{f}u}{LH^{1/6}}(\frac{2H'}{3H}+\frac{1}{u}-\frac{f'}{f})\gamma_{tt} \nonumber, \\
&&  \Theta_{xx}=\frac{\sqrt{f}u}{LH^{1/6}}(\frac{1}{u}-\frac{H'}{3H})\gamma_{xx} \nonumber, \\
&&  \Theta=\frac{\sqrt{f}u}{LH^{1/6}}(\frac{4}{u}-\frac{f'}{f}-\frac{H'}{3H}).   \nonumber
\eea Thus we obtain
\begin{equation}
T_{ij}d\vec{x}^{i}d\vec{x}^{j} = \frac{1}{8\pi G_5} (\frac{\sqrt{f}f(\pi T_0L)^2}{LH^{5/6}}(\frac{2H'}{3H}+\frac{1}{u}-\frac{f'}{f}) dt^{2}-
\frac{\sqrt{f}(\pi T_0L)^2}{LH^{-1/6}}(\frac{3}{u}-\frac{f'}{f}) d\vec{x}^{a}d\vec{x}^{a}).  \label{RBY metric}
\end{equation}
 Comparing the above stress tensor with the energy-momentum tensor of  the fluid
\begin{equation}
T_{ij}=(\varepsilon+P)u_{i}u_{j}+P \gamma_{ij},\label{SCRBY tensor}
\end{equation}
with $u^{i}=\{{\frac{\sqrt{f}\pi T_0L}{\sqrt{u} H^{1/3}},0,0,0}\}$ the normalized fluid
four-velocity,  we obtain
\begin{equation}
\varepsilon+P=\frac{1}{16\pi G}n^u(\frac{\gamma'_{tt}}{\gamma_{tt}}-\frac{\gamma'_{xx}}{\gamma_{xx}})=\frac{\sqrt{f}u_c}{\kappa^2LH^{1/6}}(\frac{H'}{H}-\frac{f'}{f})\bigg|_{u=u_c}.\label{SCRenep1}
\end{equation}
Notice the entropy density
\begin{equation}
s_c={\frac{2\pi u_c^{3/2}}{\kappa^2H^{1/2}}(1+\kappa_1)^{1/2}}\bigg|_{u=u_c}.
 \label{SCRentropyc1}
\end{equation}
On the cutoff surface, the local Hawking temperature
\begin{equation}\label{SCRtempc1}
T_c=\frac{\sqrt{u_c}H^{1/3}}{2\pi L\sqrt{f}}\frac{\kappa_1-2}{(1+\kappa_1)^{1/2}}\bigg|_{u=u_c}.
\end{equation}
So we can obtain the result according to the previous section(\ref{whata})
\begin{equation}\label{xCRchargedalx}
\frac{s_cT_c}{\varepsilon+P}=\frac{\Xi(1)}{\digamma(u_c)}
=\frac{2+\kappa_1}{2+u_c\kappa_1} \frac{(1+u_c\kappa_1)}{(1+\kappa_1)},
\end{equation} where
\be
\Xi(1)=\frac{-g^3_{xx}g'_{tt}}{\sqrt{-g}}\bigg|_{u=1}, \nonumber
\ee
\be
\digamma(u_c)=\sqrt{-g}g^{uu}\bigg(\ln {\frac{-g_{tt}}{g_{xx}}}\bigg)'\bigg|_{u=u_c}.  \nonumber
\ee
In the end,  we can find that the formulation of DC conductivity can be obtained from the combination of (\ref{SCSA}) and (\ref{xCRchargedalx})
\begin{equation}
\sigma{(u_c)}=\frac{1}{q^2(u) }\sqrt{\frac{-g}{g_{uu}g_{tt}}}g^{xx}\bigg|_{u=1} \bigg(\frac{s_cT_c}{\varepsilon+p}\bigg)^2   \label{SCRchargeu}.
\end{equation} It is identify with the equation (\ref{itis})
\be
\sigma_{DC}(u_c)=\frac{\Omega(1)}{\digamma^2(u_c)},\nonumber
\ee
The diffusion coefficient is given by
\be
\bar{D}(u_c)=\frac{\eta}{\varepsilon+P}T_c=\frac{1}{4\pi}\frac{\Xi(1)}{\digamma(u_c)}, ~~~~
\ee where $\eta(u_c)$ is given in (\ref{SCcutoffeta}). According to equations (\ref{SCRenep1}), (\ref{SCRentropyc1}), (\ref{SCRtempc1}) and the relation $\eta/s=1/4\pi$,  we can obtain
\be
\bar{D}(u_c)=\frac{1}{4\pi}\frac{2+\kappa_1}{2+u_c\kappa_1} \frac{(1+u_c\kappa_1)}{(1+\kappa_1)}.
\ee  This is  consistent with the result (\ref{SCbd1})

\section{Conclusions}
In summary, we verified the conjectured ansatz for DC conductivity and diffusion coefficient by exploring the RG flows  of charged dilatonic  and single-charged STU black holes.
The DC conductivity as well as the diffusion coefficient shows its nontrivial flow from the IR horizon to the UV boundary.
 The results indicate that black hole thermodynamics evaluated on arbitrary $u$  can provide solutions to the perturbed equation of motion for the gauge fields. We also proposed a unified
 version of retarded Green functions for shear modes in terms of black hole thermodynamic variables. A quick check shows that the formulae also work for RN-AdS black holes. The black hole thermodynamics is truly useful to describe the strongly coupled quark gluon plasma and more \cite{son1,gelaw}.

 It would be interesting to extend our work to the sound modes of charged black holes and conductivity of anisotropic and inhomogeneous holographic background, because the sound modes of hydrodynamics and conductivity of holographic lattice are very complicated and difficult to solve \cite{cgs}.  It would also be interesting to discuss the relation between the holographic RG flow and the instability problem  as noted in \cite{nojiri} and \cite{Gubser:0911}.

\renewcommand{\thesection}{\Alph{section}}
\setcounter{section}{0}

\section{Unified Retarded Green Functions}
In this appendix, we will write down the unified retarded Green function \cite{Son.Realgreenfunction} for charge dilationic black holes evaluated on the boundary theory.
We find that all the retarded Green function for the shear modes can be written in terms of black hole thermodynamics
\bea
&&G_{xt,xt}(\omega,k)=\frac{V_3}{2\kappa^2}\bigg(\frac{k^2}{i \omega-\bar{D} k^2}\bigg),\\
&&G_{xt,xz}(\omega,k)=G_{xz,xt}(\omega,k)=-\frac{V_3}{2\kappa^2}\bigg(\frac{\omega k}{i \omega-\bar{D} k^2}\bigg),\\
&&G_{xz,xz}(\omega,k)=\frac{V_3}{2\kappa^2}\bigg(\frac{\omega^2}{i \omega-\bar{D} k^2}\bigg),\\
&&G_{xt,x}(\omega,k)=G_{x,xt}(\omega,k)=-\rho\bigg(\frac{i\omega}{i \omega-\bar{D} k^2}\bigg),\\
&&G_{xz,x}(\omega,k)=G_{x,xz}(\omega,k)=\frac{\rho V_3}{8\kappa^2 P}\bigg(\frac{i\omega}{i \omega-\bar{D} k^2}\bigg),\\
&&G_{x,x}(\omega,k)=\frac{\pi\rho^2 V_3}{2 P \kappa^2 s}\bigg(\frac{i\omega}{i \omega-\bar{D} k^2}\bigg)-i\omega \sigma_{DC},
\eea
where $\kappa^2=8\pi G_5$, $V_3$ is the spatial volume along the three dimensions of the horizon, $\bar{D}$ the dimensionless diffusion constant at $u_c\rightarrow 0$, $\rho$ the charge density, $P$ the pressure, $s$ the entropy density,
and $\sigma_{DC}$ the DC conductivity at $u_c=0$.
Because both $\bar{D}$ and $\sigma_{DC}$ can be expressed by using black hole thermodynamics, the above equations should be universal expressions for charged black holes.  One can easily check the above formulae work for charge dilationic black holes,  single-charged STU black holes, and RN-AdS black holes.

For charged dilatonic black holes, the Green function $G_{x,x}(\omega,k)$ evaluated at the asymptotic AdS boundary is given by
\be
G_{x,x}(\omega,k)=\frac{Q}{\kappa^2 L} \bigg(\frac{i\omega}{i \omega-\bar{D} k^2}\bigg)-i\omega\frac{r^3_0}{2\kappa^2 L (r^2_0+Q^2)},
\ee
where $\bar{D}=\frac{1}{4\pi}\frac{r^2_0}{Q^2+r^2_0}$. One can also easily obtain  other Green functions by using the black hole thermodynamic variables given in section 2.1.

\section*{Acknowledgments}

We would like to thank Yi Ling, Yu Tian, Jian-Pin Wu, Xiao-Ning Wu and Hongbao Zhang for helpful discussions.
  This work was also supported by NSFC,
China (No.11005072 and No.11375110),  and Shanghai Rising-Star Program (No.10QA1402300).

\end{document}